%% file: chem_evol.tex
\documentclass{emulateapj}   

\pdfoutput=1

\usepackage{graphicx}
\usepackage{epsfig}
\usepackage{amsmath}
\usepackage{subfigure} 
\usepackage{amssymb}
\usepackage{apjfonts}
\usepackage{latexsym}
\usepackage{color}
\usepackage[usenames,dvipsnames,svgnames,table]{xcolor}

\shorttitle{Chemical Evolution in High Redshift Galaxies} 
\shortauthors{B.~D. Crosby et al.}

\begin{document}
\title{Tracing the Evolution of High Redshift Galaxies Using Stellar Abundances}

\author{Brian D. Crosby}
\affil{Department of Physics and Astronomy, Michigan State University, East Lansing, MI 48824, USA}
\email{crosby.bd@gmail.com}

\author{Brian W. O'Shea}
\affil{Department of Computational Mathematics, Science and Engineering, \\
Michigan State University, East Lansing, MI 48824, USA}
\affil{Department of Physics and Astronomy}
\affil{National Superconducting Cyclotron Laboratory}
\affil{JINA - Center for the Evolution of the Elements}

\author{Timothy C. Beers}
\affil{Department of Physics and JINA - Center for the Evolution of the Elements, \\
University of Notre Dame, 225 Nieuwland Science Hall, Notre Dame, IN 46556, USA}

\author{Jason Tumlinson}
\affil{Space Telescope Science Institute, 3700 San Martin Drive, Baltimore, MD, USA}

\label{firstpage}

\begin{abstract}
This paper presents the first results from a model for chemical evolution
that can be applied to N-body cosmological simulations and
quantitatively compared to measured stellar abundances from large
astronomical surveys.  This model convolves the chemical yield sets
from a range of stellar nucleosynthesis calculations (including AGB
stars, Type Ia and II supernovae, and stellar wind models) with a
user-specified stellar initial mass function (IMF) and metallicity to
calculate the time-dependent chemical evolution model for a ``simple
stellar population'' of uniform metallicity and formation time.  These
simple stellar population models are combined with a semi-analytic
model for galaxy formation and evolution that uses merger trees from
N-body cosmological simulations to track several $\alpha$- and
iron-peak 
elements for the stellar and multiphase interstellar medium
components of several thousand galaxies in the early ($z \geq 6$)
universe.  The simulated galaxy population is then quantitatively
compared to two complementary datasets of abundances in the Milky Way
stellar halo, and is capable of reproducing many of the observed
abundance trends.  The observed abundance ratio
distributions are qualitatively well matched by our model, and the
observational data is best reproduced with a Chabrier IMF, a chemically-enriched star formation efficiency of $0.2$, and a redshift of reionization of $7$.
\end{abstract}


%
%
%

\input{introduction}
\input{model}
\input{results}
\input{discussion}

\input{conclusions}
\input{acknowledgments}
\input{appendix}

\bibliographystyle{apj}
\bibliography{apj-jour,chem_evol}

\end{document}

%% file: introduction.tex
\section{Introduction}
\label{sec:introduction}

The process by which the chemical complexity in the universe was built up, from the primordial products of Big Bang nucleosynthesis to the current proliferation of elements, is an enduring question in  astrophysics.  The general picture is well understood, with Population III stars initiating chemical enrichment by expelling nucleosynthetic products into the surrounding interstellar medium (ISM), which is then subsequently incorporated into later, metal-enriched generations of stars.  This cycle of star formation, chemical enrichment within stars, and return of enriched gas to the ISM forms the foundation of galactic chemical-evolution models.  We investigate the evolution of high-redshift galaxies by building upon this basic model and connecting the nucleosynthetic yields of stellar evolution models and star formation in a cosmological context with the large-scale sets of stellar abundances observed today.

The success of $\Lambda$CDM cosmology argues for the scenario of hierarchical structure formation, with small gravitationally-bound halos repeatedly merging together to create ever-larger bound structures.  The Milky Way is the result of the smaller structures that have merged; its chemical and kinetic structure encodes the history of everything that merged to form it \citep{2013NewAR..57...80F}.  If the most metal-poor stars preserve a record of the chemical abundances of their environment at the time of their formation \citep{2002ARA&A..40..487F,2005ARA&A..43..531B}, study of the oldest, most-metal poor stars in the Milky Way, a practice termed ``Galactic Archaeology,'' can provide insight into the nature of the first stars and the evolution of high-redshift galaxies.

Observation of metal-poor stars in the Milky Way is complementary to the direct study of high-redshift galaxy formation.  Observations of high-redshift structure has progressed to the point of observing objects at $z\sim10$, providing a direct view of these early galaxies.  These impressive observations are nevertheless limited by the difficulties inherent in observing a very faint structure and deducing its characteristics.  Observations of the most metal-poor stars in the Milky Way carries a different set of advantages and disadvantages.  Spectroscopic observations of individual stars allow for precision determinations of abundances in a manner that high-redshift observations cannot accomplish, giving crucial insights into the stellar populations that preceded that star's formation.  These observations are limited by the fact that most details of the galaxy in which they form are lost during the merger process.  Analysis of the stellar halo of the Milky Way in kinematic phase space has made strides in finding stellar streams and populations from disrupted and merging structures \citep{2013NewAR..57...80F,2012ARA&A..50..251I}, but the long and complicated process of mergers that built the Milky Way has inevitably wiped away a large amount of this information.  

We have created a model that bridges these physical and temporal scales, connecting stellar nucleosynthetic yields at high redshift to the elemental-abundance patterns we currently observe in the Milky Way's oldest stellar populations.  To accomplish this we begin by developing a semi-analytic model for star formation in cosmological simulations.  We include the products of stellar nucleosynthesis within this model, and use this addition to follow the buildup of various elements as stars form, evolve, and die, as halos merge, and as diverse stellar populations are mixed.  We model each halo as multiple zones of gas and stars, hosting stellar populations of multiple ages rather than a single ``simple stellar population'' (SSP).  The stars formed in our model are similar to those currently in the Milky Way halo and dwarf galaxy populations.  By comparing the abundance patterns of our model with these observed stellar populations we can gain insight into the formation environment and history of these metal-poor stars.

This model presents a powerful new framework for interpreting observations.  Stellar evolution and nucleosynthesis models  are constantly improving, leading to a better understanding of the chemical and kinetic feedback from stellar populations.  Computing power continues to grow, enabling simulations with greater dynamic range in both spatial and mass resolution, and enabling the investigation of larger cosmological  volumes while preserving or improving spatial and temporal resolution.  Current and future observational surveys provide a wealth of stellar elemental-abundance data.  The interpretation of these data will require a model that incorporates as many relevant physical processes as possible while still maintaining computational efficiency.  We are currently able to compare to observational data from SEGUE \citep{2009AJ....137.4377Y} and the high-resolution stellar abundance measurements collected by Frebel \citep{2010AN....331..474F}.  We make quantitative comparisons between our model and observations, allowing for constraints on model parameters such as the nature of the stellar initial mass function, the efficiency with which gas forms stars, and the accuracy of theoretical stellar yields.  Ongoing and future observational campaigns such as LAMOST \citep{2012RAA....12..735D}, APOGEE \citep{2008AN....329.1018A}, Gaia-ESO \citep{2012Msngr.147...25G}, GALAH\footnote{\url{http://www.mso.anu.edu.au/galah/home.html}} \citep{2012ASPC..458..421Z}, and Gaia \citep{2012MSAIS..19..354P}, will provide vast quantities of data enabling a robust statistical comparison between model parameters and observations, bolstering our understanding of high-redshift star formation and feedback along with the nature of galactic chemical evolution in the paradigm of hierarchical structure formation.

Previous works have endeavored to investigate the formation of metal-poor stars.  The semi-analytic models of \citet{2006ApJ...638..585F} construct merger histories for several galaxies similar to the Milky Way using a method based on the extended Press-Schechter formalism \citep{1993MNRAS.262..627L}, and follow the abundances of alpha group elements in addition to iron to track the chemical distribution of the galactic halos at their satellites.  \citet{2010MNRAS.401L...5S} combine \emph{N}-body simulations with semi-analytic models for star formation in a single-phase interstellar medium to track metallicity via iron abundance and model the spatial metallicity distribution of stars in the Milky Way.  \citet{2010ApJ...708.1398T} utilizes semi-analytic chemodynamic modeling of high-redshift star formation to probe the degree to which the Milky Way halo stars reflect the star formation environment present in halo progenitor galaxies before, during, and after the epoch of reionization.  The abundance and spatial distribution of metal-poor stars in the galaxy are used to probe the Population III initial mass function by \citet{2011ApJ...736...73K}, who also construct a merger history using the extended Press-Schechter formalism and expand the number of elemental abundances that they track to seven, but focus on yields from supernovae rather than the entire ensemble of manners in which stars chemically and dynamically enrich the surrounding medium.  \citet{2011ApJ...729...16K} couple semi-analytic models of star formation to \emph{N}-body cosmological simulations of a Milky Way-like disc galaxy, tracking the abundances of 13 elements, but due to their high particle mass of $1.0-3.8\times10^6~$M$_{\odot}$ and late  initial redshift of 24, these simulations do not resolve the formation of high-redshift and low-mass progenitor galaxies.  

This model overcomes many of the challenges that have limited previous work on this subject by combining robust, physically motived semi-analytic models of star formation and stellar feedback with well-resolved $N$-body cosmological simulations, and a suite of yields from stellar evolution simulations.  Additionally, this model has the capability to make statistically significant, quantitative comparisons to current and future observations rather than the qualitative analysis that pervades previous works.  These initial models can be readily extended as more complete and self-consistent stellar yields become available, and the data from this model can be naturally paired with powerful statistical tools such as Gaussian Multiprocess emulation coupled Markov Chain Monte Carlo tools and ANOVA decomposition \citep{2012ApJ...760..112G, 2014ApJ...787...20G}, enabling the rapid exploration and evaluation of the parameter space of these models.

This is the second paper in a series.  In \citet{2013ApJ...773..108C} (hereafter Paper I), we presented a semi-analytic model for calculating the star formation history of all halos in a cosmological simulation  across the full temporal extent of the simulation, thus allowing us to identify halos where Population III or metal-enriched star formation are taking place.  This paper investigates the chemical evolution of a population of high-redshift galaxies through the use of a semi-analytic chemical enrichment model that was built on top of the model described in Paper I.  Synthetic stellar populations are created for every halo in the simulation, and a stellar feedback model is implemented to connect the evolution of the stellar population to the ongoing star formation in each halo.  With this model, chemical evolution histories for the stellar and gaseous components of all of the star-forming halos in the simulation are created, and can then be compared directly to observations of metal-poor stars.  The outline of this paper is as follows:  A brief review of the simulations used, the star formation model, and the chemical-evolution model are given in Section \ref{sec:model}.  Our results are presented in Section \ref{sec:results}, and compared to observational data and other theoretical work in Section \ref{sec:discussion}, where we also include a discussion of the limitations of this study.  Finally, we present a summary of our conclusions in Section \ref{sec:conclusions}.

%% file: model.tex
\section{Model Description}
\label{sec:model}

\subsection{Overview}
The model employed in this work is an extension and modification of the star formation model presented in Paper I \citep[i.e.,][]{2013ApJ...773..108C}.  This model tracks the chemical evolution of halos across time by following 10 chemical species:  H, C, N, O, Mg, Ca, Ti, Fe, Co, and Zn, in both the stellar and interstellar medium (ISM) components of every halo.  A more detailed treatment of the ISM than was implemented in Paper I is required to accurately model the star formation and feedback processes.  Paper I treated the gas and metal quantities in a halo as a single zone, in which material from accretion and stellar feedback was mixed instantaneously throughout the entire halo, which does not accurately reflect the dispersal and recycling of material between the ISM and the stellar population.  The ISM is now treated as a multiphase gas with a central region of dense, cold gas that is capable of forming stars and a hot, diffuse region exterior to the star-forming central region that is incapable of forming stars.  This two-phase medium provides a simple framework in which to investigate the bulk properties of the ISM in halos \citep{2005ARA&A..43..337C}.  Throughout the remainder of this paper, this exterior region of warm, diffuse gas will be referred to as the ``reservoir'' of the halo, and the cold, dense, star-forming region will be referred to as the ``central'' region.  Gas and chemical species are moved between these two ISM regions and the stellar component in each halo.  Material accreted from the intergalactic medium is deposited in the reservoir, and gas in the reservoir can cool and transition to the central region.  Gas in the central region is available to condense and form stars, and feedback from the stellar population returns enriched material to this region.  Kinetic feedback from supernovae (SNe) will move enriched material from the stellar population to the central region, as well as to the reservoir.  If the feedback from SNe is sufficiently powerful, it will eject material to the intergalactic medium (IGM), permanently removing it from the halo.  The onset of reionization generally suppresses star formation in halos, with a greater attenuation occurring in low mass halos, and a lesser impact on higher mass halos.

The Population III star formation model used in this work is presented in Paper I, and a brief summary is given here in Section \ref{sec:pop III sf}.  The interested in reader is encouraged to see \citet{2013ApJ...773..108C} for a detailed discussion.  The treatment of the multi-phase interstellar medium and chemically-enriched star formation implemented in this work is discussed in Section \ref{sec:chemically enriched sf}.

A summary of the parameters of the model that were tested, their fiducial values, and the ranges investigated is given in Table \ref{table:parameters}.

\begin{table*}[htdp]
\caption{Model parameters with their fiducial values, the range tested, and a brief description.}
\begin{center}
\begin{tabular}{ l c c l }
\multicolumn{4}{c}{Model Parameters}	\\
\hline
\hline
Parameter	&	Fiducial Value	&	Range	&	Description	\\
\hline
$\mathcal{E}$	&	0.04		&	0.008-0.2					&	Star formation efficiency \\
$f_{\rm{esc}}^{\rm{LW}}$	&	1		&	0.01, 0.1, 1		&	LW photon escape fraction	\\
IMF				&	Salpeter	&	Salpeter, Kroupa, Chabrier	&	Chemically-enriched stellar IMF 	\\
$z_{\rm{reion}}$	&	8		&	8, 7, 6.5, none				&	Redshift of reionization \\
\hline
\end{tabular}
\end{center}
\label{table:parameters}
\end{table*}

The simulations used as a basis for our model were carried out using the publicly available Enzo adaptive mesh refinement + \emph{N}-body code\footnote{\url{http://enzo-project.org}} \citep{2014ApJS..211...19B}, and are the same simulations that were used in Paper I.  Four simulations were run -- two with a comoving box size of $3.5~h^{-1}$Mpc and two with a comoving box size of $7.0~h^{-1}$Mpc.  We used two different simulation volumes and two random realizations per chosen volume to give some idea of the impact of cosmic variance as well as mass and spatial resolution on our results.  We use the WMAP 7 best-fit cosmological model \citep{2011ApJS..192...18K}, with $\Omega_{\Lambda}=0.7274$, $\Omega_{M}=0.2726$, $\Omega_{B}=0.0456$, $\sigma_8=0.809$, $n_{s}=0.963$, and $h=0.704$ in units of $100~$km s$^{-1}$Mpc$^{-1}$, with the variables having their usual definitions.  All simulations are cubic and have 1024 grid cells per edge and $1024^3$ dark matter particles, giving cell dimensions of $6.8~h^{-1}$ comoving kpc on a side, a dark matter particle mass of $2.86\times10^4~$M$_\odot$, a mean baryonic mass per cell of $5.74\times10^3~$M$_\odot$, and a total mass of $3.7\times10^{13}~$M$_\odot$ for the $7.0~h^{-1}$Mpc boxes.  The $3.5~h^{-1}$Mpc boxes have cell dimensions of $3.4~h^{-1}$ comoving kpc on a side, a dark matter particle mass of $3.57\times10^3~$M$_\odot$, a mean baryonic mass per cell of $718~$M$_\odot$, and a total mass of $4.6\times10^{12}~$M$_\odot$.  These volumes contain enough mass to form a galaxy with mass similar to that of the Milky Way.  In the current epoch, approximately half of all galaxies reside in groups.  At high redshift, the progenitors of the Milky Way and other Local Group galaxies are decidedly average, though possibly somewhat more strongly clustered than field galaxies at the equivalent redshift \citep{2013ApJ...773..105C}.  Thus, the Milky Way progenitor population is comparable to the structures in the simulations used in this work, and a volume containing the progenitors of a Milky Way-like galaxy is statistically comparable to a randomly chosen cosmological volume of similar size.

The simulations were initialized at $z=99$ using the MUSIC cosmological initial conditions generator \citep{2011MNRAS.415.2101H}, with a second-order Lagrangian perturbation theory method and separate transfer functions for dark matter and baryons.  A second-order Lagrangian perturbation method is necessary to obtain converged halo mass functions at such early times and high redshifts as the start of Population III star formation \citep{2006MNRAS.373..369C}.  Each of the sets of initial conditions were generated using a different random seed.  The simulations were run with Enzo's unigrid (non-adaptive mesh refinement) mode with adiabatic hydrodynamics, from $z=99$ to $z=6$.  Data is output at integer redshifts until $z=14$, at which point the elapsed time between integer redshifts would exceed the timescale for star formation.  After $z=14$, data is output every $11~$Myr.  The simulation is stopped at $z=6$ to prevent modes on the order of the size of the simulation volume from entering the non-linear regime.  We note that extensive physics (e.g., radiative cooling, star formation and feedback) is unnecessary in these simulations, as they are simply being used as the source of merger trees for our semi-analytic models.

Dark matter halos for all data outputs in the simulations were identified using the Friends-of-Friends \citep{1985ApJS...57..241E} halo finder implemented in the yt analysis toolkit\footnote{\url{http://yt-project.org}} \citep{2011ApJS..192....9T}, with a linking length of $0.2$ times the mean interparticle spacing.  Halo merger trees were then created to show the assembly history of these dark matter halos based on particle membership.

\subsection{Population III Star Formation}
\label{sec:pop III sf}
Population III stars form in chemically-pristine halos that cool via H$_2$ to a temperature and density at which the core is unstable to gravitational collapse \citep{2002Sci...295...93A,2007ApJ...654...66O}.  In our model, a halo is deemed to be capable of hosting Population III star formation if the cooling timescale is less than the local Hubble time.  In the absence of a H$_2$ phototdissociating background, this can be cast as a minimum mass threshold that depends only on the redshift, $z$,
 
\begin{equation}
M_{\rm{min,Hubble}}=5.87\times 10^4 \left(\frac{1+z}{31}\right)^{-2.074}\rm{M}_\odot .
\label{m_min_hubble}
\end{equation}

This mass threshold will be modified in the presence of other stars, as radiation in the Lyman-Werner (LW) band (11.18-13.6 eV) is capable of photodissociating H$_2$, suppressing cooling.  The minimum halo mass threshold for Population III star formation in the presence of LW background radiation becomes 
\begin{equation}
M_{\rm{min,LW}}=1.91\times 10^6 J_{21}^{0.457}\left(\frac{1+z}{31}\right)^{-2.186}\rm{M}_\odot,
\label{m_min_LW}
\end{equation}
where $J_{21}$ is the proper LW flux.  $J_{21}$ is defined from the comoving LW  photon number density, $n_{\rm{LW}}$, in Mpc$^{-3}$, as
\begin{equation}
J_{21}=1.6\times10^{-65}n_{\rm{LW}}\left(\frac{1+z}{31}\right)^3~\rm{erg~s^{-1}~cm^{-2}~Hz^{-1}~sr^{-1}}.
\label{j21}
\end{equation}
The effective $J_{21}$ seen by each pristine halo is modified to account for H$_2$ self-shielding following \citet{2011MNRAS.418..838W}.  The minimum halo mass threshold for Population III star formation is then taken to the most stringent of Equations \ref{m_min_hubble} and \ref{m_min_LW}, 

\begin{equation}
M_{\rm{min}}=\rm{max}
	\begin{cases}
	5.87\times 10^4 \left(\frac{1+z}{31}\right)^{-2.074}\rm{M}_\odot	\\
	1.91\times 10^6 J_{21}^{0.457}\left(\frac{1+z}{31}\right)^{-2.186}\rm{M}_\odot
	\end{cases}.
	\label{minmass}
\end{equation}

Any halo that is chemically-pristine and more massive than the mass threshold for Population III star formation is assumed to form a star.  The halo is tagged as chemically enriched, and it and all of its descendants are no longer capable of forming a Population III star.  When a Population III star is formed, a delay time prior to the start of chemically-enriched star formation is determined based on the assumed Population III stellar lifetime.  The delay time is scaled inversely with the halo mass to account for gas blown out of the halo by a Type II supernova (SNII), which is assumed to be the end of all Population III stars in this model.  A full description of the Population III star formation model is presented in Paper I.

\subsection{Chemically Enriched Star Formation}
\label{sec:chemically enriched sf}
Any halo that contains particles that had previously been in a halo that formed stars is deemed incapable of forming a Population III star due to metal pollution, and star formation in these chemically-enriched halos is treated differently.  Chemically-enriched star formation is modeled as a continuous process, as opposed to Population III star formation, which is discrete and a function of the mass of gas available in the halo \citep{2010ApJ...724..687L}.  We note that chemically-enriched star formation is treated somewhat differently than in Paper I.  The baryons in a given halo are modeled as three interacting populations: a reservoir of gas in the outskirts of the halo that is hot and diffuse, and thus does not form stars; a mass of gas in the center of the halo that is cold, dense, and star-forming; and a mass of baryons that are currently locked up in stars.  A set of three differential equations governs star formation and gas transport between the reservoir and central regions of a chemically-enriched halo,
\begin{equation}
\frac{dM_{\rm{res}}}{dt} = -\frac{M_{\rm{res}}}{\tau_{\rm{cool}}} + \Lambda \frac{r_{\rm{c}}}{M_{\rm{c}}}\frac{1}{\Delta t} - \Lambda \frac{r_{\rm{halo}}}{M_{\rm{halo}}}\frac{1}{\Delta t}~,
\label{dmresdt}
\end{equation}
\begin{equation}
\frac{dM_{\rm{c}}}{dt} = \frac{M_{\rm{res}}}{\tau_{\rm{cool}}} - \Lambda \frac{r_{\rm{c}}}{M_{\rm{c}}}\frac{1}{\Delta t} - \frac{\mathcal{E}}{\tau} M_{\rm{c}} + \frac{M_{\rm{eject}}}{\Delta t}~,  
\label{dmsfdt}
\end{equation}
\begin{equation}
\frac{dM_\star}{dt} = \frac{\mathcal{E}}{\tau} M_{\rm{c}}~.
\label{dmstardt}
\end{equation}

Equation \ref{dmresdt} governs the rate of change of the mass of gas in the reservoir, Equation \ref{dmsfdt} governs the rate of change of the mass of gas in the central region, and Equation \ref{dmstardt} governs the star formation rate.  In Equation \ref{dmresdt}, the first term represents gas that cools from the reservoir and condenses into the central region, and $\tau_{\rm{cool}}$ is the cooling time of the gas, which is calculated using the Grackle chemistry and cooling library\footnote{\url{https://grackle.readthedocs.org/}} \citep{2014ApJS..211...19B,2014ApJS..210...14K}.  The second term represents gas ejected from the supernovae in the central region that enters the reservoir, where $\Delta t$ is the integration timestep and $\Lambda$ is a factor encapsulating information about the supernovae that occurred during this integration timestep (see Equation~\ref{lambda} in Section \ref{sec:halo_ejection}).  The third term is similar to the second, but represents gas ejected completely from the halo via supernovae.  In Equation \ref{dmsfdt}, the first two terms are the same as the first two terms in Equation \ref{dmresdt} but with the opposite signs, to reflect gas cooling into the central region and gas ejected from it via supernovae, respectively.  The third term models the conversion of gas in the central region into stars, where $\mathcal{E}$ is the dimensionless star-formation efficiency (SFE) and $\tau$ is the characteristic star formation time, taken to have a constant value of $\tau = 10^8$ years.  The fourth term represents gas expelled by the stars to the interstellar medium.  Equation \ref{dmstardt} is equivalent to the third term in Equation \ref{dmsfdt}, and creates a mass of stars from the available gas in the central region.  Equation \ref{dmstardt} forms a mass of stars rather than individual stars, allowing for different stellar initial mass functions (IMFs) to be applied.  The time interval between each simulation data output is traversed in 100 integration timesteps, advancing Equations \ref{dmresdt}-\ref{dmstardt} forward in time.  The age distribution of the stellar component of each halo is tracked in 100 linearly spaced age bins, spanning the time that the first star formed in the simulation to the end of the current simulation data output.  As time passes, the stellar content is advanced through the age bins, enabling each stellar age bin in every halo to be evolved as a SSP.  Every stellar age bin returns gas and enriched material to the halo interstellar medium (ISM) at each integration timestep.  Additionally, the expected number of supernovae is determined and material is ejected from the halo to the intergalactic medium (IGM).  The halo ejection model is presented in Section \ref{sec:halo_ejection}.  New stars that form in a halo are formed with a fraction of metals equal to that of the ISM at the time of formation.  A complete description of the chemical-evolution model is given in Section \ref{sec:chemical_evolution}.  

The onset of reionization attenuates star formation in halos, and has a more pronounced effect on smaller halos than on larger ones.  We model reionization by drawing inspiration from \citet{2010ApJ...708.1398T}.  After the onset of reionization, star formation in halos with a circular velocity less than $30$ km s$^{-1}$ is completely suppressed.  Halos with a circular velocity in excess of $50$ km s$^{-1}$ experience no suppression, and those with circular velocities falling within this range experience suppression in SFE that varies linearly with circular velocity.

As halos merge between simulation data outputs, all attributes of parent halos are inherited by the child halo in proportion to the fraction of the parent halo mass that is inherited by the child halo.  Child halos generally inherit either all or the majority of the baryonic content of the parent halos that merge to form it.  This includes inheriting the mass of gas available to form stars in the halo, the mass of all chemical species in the ISM, the mass of all chemical species in the stellar components, and the stellar population along with its age distribution.  The age distribution is remapped at each subsequent simulation data output.  This remapping decreases the time resolution as the population ages, and is done to prevent excessive computational memory usage.  This remapping produces a maximum stellar age bin size of $3.72$ Myr, which is larger than only the smallest stellar age bin in the tabulated data of material returned to the ISM by asymptotic giant branch (AGB) stars, resulting a temporal resolution of the chemical and kinetic feedback model being limited primarily by the time resolution of the available stellar-feedback data.

\subsection{Gas and Metal Ejection}
\label{sec:halo_ejection}
The prescription for ejection of material via supernovae from a halo to the IGM is based several quantities --  the number of supernovae that occurred; the mass of gas ejected to the ISM by supernovae, $M_{\rm{gas}}^{\rm{ISM}}$; the mass of species $Z$ ejected to the ISM by supernovae, $M_{\rm{Z}}^{\rm{ISM}}$; and the virial parameters of the halo, specifically the mass ($M_{\rm{vir}}$) and radius ($r_{\rm{vir}}$).  

The first step of this process is to determine the mass of gas ejected from the halo to the IGM as a result of supernovae, $M_{\rm{lost}}$.  This is accomplished following \citet{2010ApJ...708.1398T}, by comparing the energy imparted to the halo gas by all of the supernovae that exploded during the current timestep to the kinetic energy of gas moving at the halo escape velocity, $v_{\rm{esc}}$.  
\begin{equation}
E_{\rm{SNe}}=E_{\rm{wind}}=\frac{1}{2}M_{\rm{lost}}v_{\rm{esc}}^2
\label{energetics}
\end{equation}
Solving Equation \ref{energetics} for $M_{\rm{lost}}$ and using the definition of the escape velocity approximating the halo as spherical, with mass $M_{\rm{vir}}$ and radius $r_{\rm{vir}}$ allows for the calculation of the $M_{\rm{lost}}$ in terms of the supernova energetics and the halo physical properties,
\begin{equation}
M_{\rm{lost}} = \frac{E_{\rm{SNe}}r_{\rm{vir}}}{GM_{\rm{vir}}}~.
\label{m_gas}
\end{equation}
The energy imparted to the wind by supernovae can be parameterized as $E_{\rm{SNe}} = N_{\rm{SNe}}\epsilon_{\rm{SNe}}E_{51}$, where $N_{\rm{SNe}}$ is the number of supernovae that occurred during the current timestep, $\epsilon_{\rm{SNe}}$ is the efficiency with which the supernova energy is converted to the kinetic energy of the gas, and $E_{51}$ is the energy of a single supernova in units of $10^{51}~$ergs.  A fiducial value of $\epsilon_{\rm{SNe}}=0.0015$ is adopted following \citet{2010ApJ...708.1398T}, which makes the assumptions that $5\%$ of the total supernova energy is kinetic, and that of this $3\%$ is transferred to the ejected material.  In both Paper I and \citet{2011ApJ...736...73K}, the evolution of the simulation and metallicity distributions functions are largely independent of the precise value of the  $\epsilon_{\rm{SNe}}$.  As such,  $\epsilon_{\rm{SNe}}$ will not be varied in this work.  Using this parameterization along with Equation \ref{m_gas} allows for the calculation of $M_{\rm{lost}}$ as
\begin{equation}
M_{\rm{lost}} = 7.792\times10^8 \frac{N_{\rm{SNe}}\epsilon_{\rm{SNe}}E_{51}r_{\rm{vir}}}{GM_{\rm{vir}}}~\rm{M}_\odot,
\label{m_lost}
\end{equation}
where $M_{\rm{vir}}$ is the halo virial mass in units of M$_\odot$, $r_{\rm{vir}}$ is the halo virial radius in proper Mpc, and $G$ is the gravitational constant is in CGS units.  We also define the quantity 
\begin{equation}
\Lambda = 7.792\times10^8 \frac{N_{\rm{SNe}}\epsilon_{\rm{SNe}}E_{51}}{G}~,
\label{lambda}
\end{equation}
for use in Equations \ref{dmresdt}-\ref{dmstardt} to calculate the mass of gas that is ejected via supernovae from the central region to the reservoir region, as well as out of the halo entirely.

\subsection{Chemical Evolution}
\label{sec:chemical_evolution}
This model tracks the abundances of 10 elements in both the central and reservoir regions of the ISM, as well as the elemental masses in the stellar component of every halo, in addition to the total mass of gas available for star formation and in the reservoir.  The elements H, C, N, O, Mg, Ca, Ti, Fe, Co, and Zn are followed.  Tabulated yields from stellar evolution and nucleosynthesis simulations are used to determine the masses of each of these elements that are ejected from stars to the ISM, as well as the total ejected mass of gas.

Chemical enrichment in a Population III star-forming halo is treated differently than that in a chemically-enriched halo due to the different manners in which star formation is modeled in these different environments.  All Population III stars are assumed to end their lives as Type II supernovae, at which point they return enriched material to the halo ISM, using yields from \citet{2002ApJ...567..532H}.  The initial mass of each Population III star is randomly selected with equal probability from all masses for which yields are available, ranging from $30$M$_\odot$ to $100$M$_\odot$.  A factor representing the Population III stellar multiplicity in each halo is adopted, and has a user-defined fiducial value of $1.2$, motivated by the findings of \citet{2009Sci...325..601T} which show fragmentation in the pre-stellar cloud, suggesting the possibility of the formation of Population III binary star systems.  Gas and enriched material is returned to the halo in a quantity equal to the tabulated yields multiplied by the Population III multiplicity factor, treating the possibility of Population III binary star systems in a statistical manner.

The movement of chemical species between the central, reservoir, and stellar components of a halo is governed by a set of differential equations similar to Equations \ref{dmresdt}-\ref{dmstardt},
\begin{equation}
\frac{dM_{\rm{res}}^{\rm{Z}}}{dt} = \frac{M_{\rm{to~res}}^{\rm{Z}}}{\Delta t} - \frac{M_{\rm{lost}}^{\rm{Z}}}{\Delta t} - \frac{M_{\rm{res}}^{\rm{Z}}}{\tau_{\rm{cool}}} ~,
\label{dmzresdt}
\end{equation}
\begin{equation}
\frac{dM_{\rm{c}}^{\rm{Z}}}{dt} = \frac{M_{\rm{res}}^{\rm{Z}}}{\tau_{\rm{cool}}} - \frac{M_{\rm{c}}^{\rm{Z}}}{\tau} + \frac{M_{\rm{eject}}^{\rm{Z}}}{\Delta t} - \frac{M_{\rm{to~res}}^{\rm{Z}}}{\Delta t} ~,
\label{dmzsfdt}
\end{equation}
\begin{equation}
\frac{dM_\star^{\rm{Z}}}{dt} = \frac{M_{\rm{c}}^{\rm{Z}}}{\tau} ~,
\label{dmzstardt}
\end{equation}
where Equation \ref{dmzresdt} describes the time evolution of the mass of species Z in the reservoir gas that is not forming stars, Equation \ref{dmzsfdt} describes the same for the central region, and Equation \ref{dmzstardt} describes the evolution of the mass of species Z in the stellar component.  The superscript Z denotes that the quantity is the mass of a chemical species, and is used to differentiate from the equation governing the evolution of the gas in Equations \ref{dmresdt}-\ref{dmstardt}.  $M_{\rm{to~res}}^{\rm{Z}}$ is the mass of species Z that moves from the central region to the reservoir region in a given integration timestep, and $M_{\rm{lost}}^{\rm{Z}}$ is the mass ejected from the halo to the IGM.  These two terms are determined by the number of supernovae that occur and the mass of gas and each chemical species ejected by supernovae in the halo during a given integration timestep.  Gas and metals are instantaneously mixed within each ISM component.
  
The kinetic energy imparted to supernova products should give rise to preferential ejection of this material from the halo \citep{2010ApJ...708.1398T}.  The mass of gas ejected from the halo to the IGM by supernovae is determined by Equation \ref{m_lost}.  This quantity is used in conjunction with the mass of gas ejected to the ISM by supernovae to determine both the mass of each element ejected from the central region to the reservoir region, and from the halo entirely.  Looking first at $M_{\rm{to~res}}^{\rm{Z}}$, a mass of element Z produced in supernovae will reach the reservoir in proportion to the ratio of the mass of gas ejected to the reservoir to the total mass of gas ejected by supernovae, 
\begin{equation}
M_{\rm{to~res}}^{\rm{Z}} = M_{\rm{eject}}^{\rm{Z}}\frac{M_{\rm{to~res}}}{M_{\rm{eject}}} = 
 M_{\rm{eject}}^{\rm{Z}}\frac{\Lambda r_{\rm{c}}/M_{\rm{c}}}{M_{\rm{eject}}}~.
\label{mztores}
\end{equation}
This method is similarly used to determine the mass of each species ejected from the reservoir to the IGM,
\begin{equation}
M_{\rm{lost}}^{\rm{Z}} = M_{\rm{eject}}^{\rm{Z}}\frac{\Lambda r_{\rm{vir}}/M_{\rm{vir}}}{M_{\rm{eject}}}~.
\label{mztoigm}
\end{equation}

Chemical feedback from stars in chemically-enriched halos is modeled differently than in Population III star-forming halos due to the differences in star formation methods.  Each stellar age bin in a halo is treated as a SSP of uniform metallicity and identical star formation time that will eject gas and chemically-enriched material as it ages.  Time-dependent yield tables were created by convolving the yields from stellar evolution models with weightings from an adopted stellar initial mass function (IMF).  Different version of these tables were created for Salpeter \citep{1955ApJ...121..161S}, Chabrier \citep{2003PASP..115..763C}, and Kroupa \citep{2002Sci...295...82K} IMFs at various metallicities.  The three IMFs have functional forms

\begin{eqnarray}
\frac{dN}{dm}=&\Phi_{\rm{Salpeter}}=&0.154m^{-2.35}\\
&\Phi_{\rm{Kroupa}}=&\left\{ \begin{array}{lc}
0.56m^{-1.3}& m\le0.5\rm{M}_{\odot}\\ 
0.3m^{-2.2}&0.5\rm{M}_{\odot} < m \le1\rm{M}_{\odot}\\ 
0.3m^{-2.7}& m> 1\rm{M}_{\odot}
\end{array}\right.\\
&\Phi_{\rm{Chabrier}}=& \left\{\begin{array}{lc} 
\frac{0.799}{m} e^{-(\log m/m_{\rm c})^2/2\sigma^2}&  m \leq 1 \rm{M}_{\odot} \\ 
0.223 m^{-2.3} & m > 1 \rm{M}_{\odot}
\end{array}\right.\\
\end{eqnarray}

\begin{figure}[htbp] 
   \centering
   \includegraphics[width=.45\textwidth, clip=true]{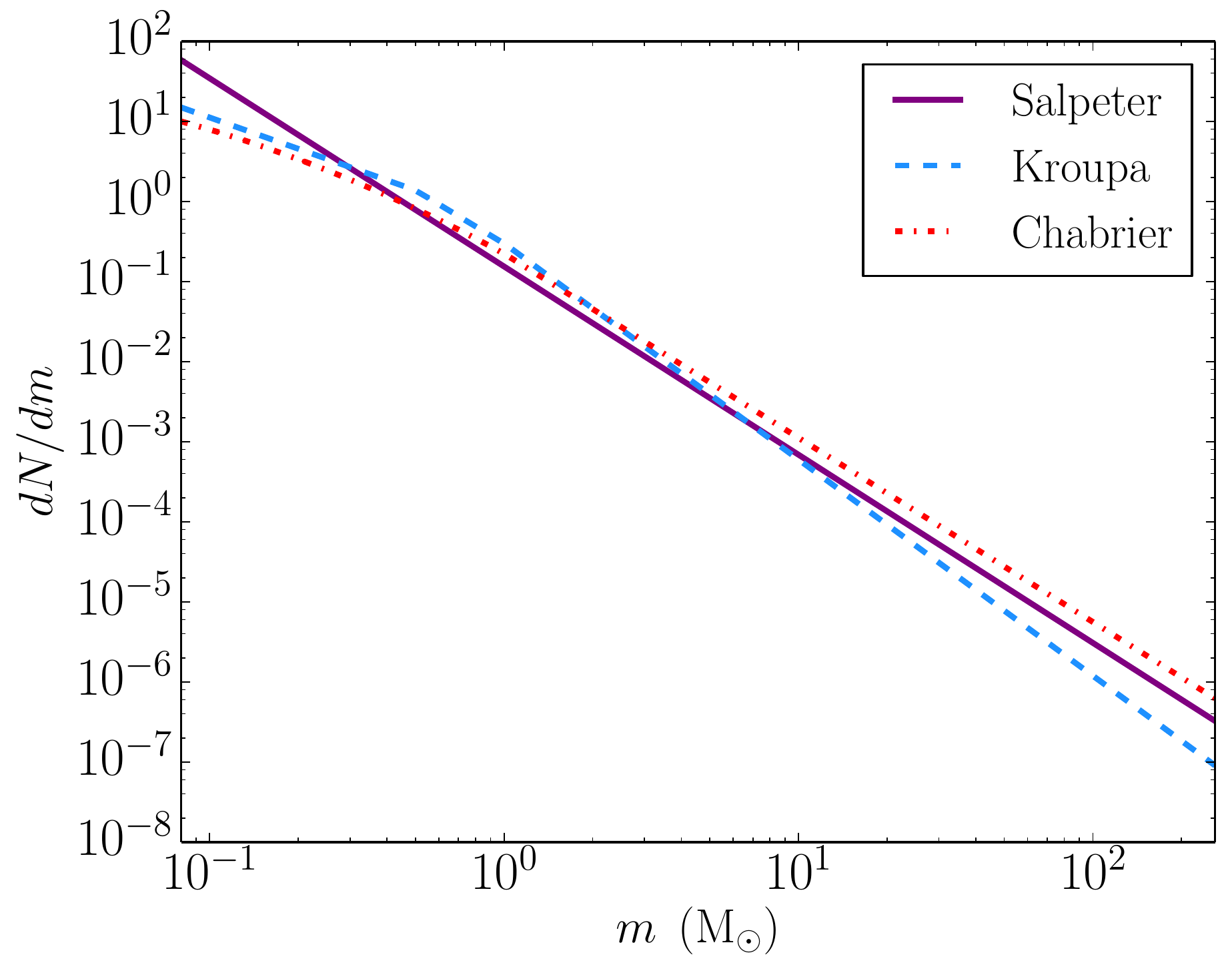} 
   \caption{The three IMFs considered in this work are Salpeter (violet, solid line), Kroupa (blue, dashed line), and Chabrier (red, dot-dashed line), over a mass range of $0.08$~M$_\odot$ to $260$~M$_\odot$.  The integrated area under each of the curves is the same.  The Salpeter IMF emphasizes low-mass stars, the Kroupa IMF emphasizes intermediate-mass stars, and the Chabrier IMF is by the far the most top-heavy of the three, emphasizing high-mass stars.}
   \label{fig:imf_plot}
\end{figure}

In the Chabrier IMF $m_{\rm{c}}$ is the characteristic mass and takes a value of $0.079$~M$_\odot$ and the dispersion $\sigma=0.69$ \citep{1955ApJ...121..161S,2002Sci...295...82K,2003PASP..115..763C}.  The IMFs are all considered to be applicable in a mass range of $0.08$~M$_\odot$ to $260$~M$_\odot$ and are shown in Figure~\ref{fig:imf_plot}.  Tables were created following the yields from Type Ia supernovae (SNIa), SNII, and AGB stars.  Additional tables were created to give the rates of SNIa and SNII as a function of time since the formation of the SSP.  These tables were created for all three IMFs and for four different metallicities.  The metallicity of the gas in the central star-forming region of each halo is examined at each integration timestep, and the appropriate yields are used.  Details of the creation of the yield tables are presented in Appendix \ref{app:yields}.  

At every integration timestep, the stellar metallicity in each halo is used to determine which yield set will be used for the stellar feedback.  All of the stellar age bins use these tables to eject gas and chemically-enriched material into the ISM.  The number of supernovae predicted to occur in the halo is determined, and the mass of gas thus ejected to the IGM is calculated using Equation \ref{m_lost}.  This is compared to the mass of gas ejected by supernovae to the ISM, and either Equation \ref{mztores} or \ref{mztoigm} is used to update the mass of all chemical species in the ISM, setting the composition of the next stars to form in this halo.

%% file: results.tex
\section{Results}
\label{sec:results}
This model tracks the chemical evolution of 10 elements in all of the halos in a simulation:  H, C, N, O, Mg, Ca, Ti, Fe, Co, and Zn.  Comparison of the synthesized stellar metallicity distributions to observational data allows for the evaluation of our model, and to place constraints on the parameters we have chosen to investigate.  We compare our results to the limited set of elemental-abundance ratios ([Mg/Fe], [C/Fe]) determined for the large sample of low-resolution stellar spectra from the SEGUE database \citep{2009AJ....137.4377Y}, and to the much smaller sample of metal-poor stars with available high-resolution spectroscopic elemental-abundance ratios assembled by \citet{2010AN....331..474F}. All elements tracked in this model are represented in the Frebel dataset.

The chemical evolution of the stellar populations of all halos in the [C/Fe]-[Fe/H]\footnote{We adopt the standard convention [X/Y]$=\log_{10}(N_{\rm{X}}/N_{\rm{Y}})-\log_{10}(N_{\rm{X}}/N_{\rm{Y}})_\odot$.} space are shown in Figure \ref{fig:chabrier_CFe}, and in [Mg/Fe]-[Fe/H] space in Figure \ref{fig:chabrier_MgFe} with observational data from SEGUE \citep{2009AJ....137.4377Y} and \citet{2010AN....331..474F} overplotted in red and yellow, respectively.  Data from these sources has been binned in $0.25$ dex increments in [Fe/H], with the stellar number weighted mean plotted as a continuous line, the $68\%$ confidence interval shown as thick lines, and the minimum and maximum extents are shown as thin lines.  These figures show the abundance distributions of the stellar mass produced by our model at $z=6$.  The central panel of these figures shows the mass-weighted abundance distribution of stellar material as a shaded blue region (with dark blue indicating areas with larger fraction of the total stellar mass, and light blue representing areas with a very small fraction of the total stellar mass), while the top and right panels are histograms of the abundance distribution of stellar mass for a single quantity of interest.  It should be noted that the peaks of the histograms showing the distribution in individual abundances are not necessarily at the same values as the peak in the central panel, as the peak of the central panel corresponds to the most common \textit{pair} of abundances.  For example the most common combination of [Fe/H] and [Mg/Fe] might have an [Fe/H] value that is not the same as the most common value when the distribution of stellar mass in only [Fe/H] is considered.

\subsection{Comparison to Observations}

Quantitative comparison is made to the observational data by calculating the implausibility value,
\begin{equation}
\mathrm{I} = \frac{\sum_{\mathrm{bins}}{\left(\langle[\mathrm{X/Y}]\rangle_{\mathrm{obs}} - \langle[\mathrm{X/Y}]\rangle_{\mathrm{sim}}\right)^2}}    {\sum_{\mathrm{bins}}{\left(\sigma_{\rm{obs}}^2 + \sigma_{\mathrm{sim}}^2 \right)}}~,
\label{implausibility}
\end{equation}
and a joint probability metric,
\begin{equation}
\begin{aligned} 
\mathrm{J}=\Big( \sum_{\rm{all~bins}} P_{i,\rm{sim}} P_{i,\rm{obs}} \Big) 
\Bigg[n_{\rm{overlap}}
\Big( \sum_{\rm{overlap}} P_{j,\rm{sim}} + \sum_{\rm{overlap}} P_{j,\rm{obs}} \Big) \\
- \Big( n_{\rm{lone,sim}} \sum_{\rm{lone}} P_{k,\rm{sim}} + n_{\rm{lone,obs}}\sum_{\rm{lone}} P_{l,\rm{obs}} \Big) \Bigg]\frac{1}{n_{\rm{dists}}}
\end{aligned}
\label{joint}
\end{equation}
where the simulated and observed data have been binned into square, half dex increments in [X/Fe]-[Fe/H] space.  In Equation \ref{implausibility}, $\langle[\mathrm{X/Y}]\rangle_{\mathrm{obs}}$ and $\langle[\mathrm{X/Y}]\rangle_{\mathrm{sim}}$ are the observed and simulated weighted means in each bin, and $\sigma_{\mathrm{obs}}$ and $\sigma_{\mathrm{sim}}$ are the associated standard deviation in each bin for each dataset.  In Equation \ref{joint}, $n_{\rm{dists}}$ is the number of distributions being compared, $P_{i,\rm{sim}}$ and $P_{i,\rm{obs}}$ are the simulated and observed probability densities in bin $i$, $n_{\rm{overlap}}$ is the number of bins in which the simulated and observed elemental-abundance distributions overlap, and $n_{\rm{lone,sim}}$ and $n_{\rm{lone,obs}}$ are the number of non-overlapping bins in the simulated and observed distributions, respectively.

Equations \ref{implausibility} and \ref{joint} were used to evaluate the [C/Fe]-[Fe/H] and [Mg/Fe]-[Fe/H] distributions from the SEGUE dataset, and the [X/Fe]-[Fe/H] distributions for all abundances in the Frebel dataset. The implausibility is a metric by which to assess the agreement of the simulated and observed elemental-abundance distributions while accounting for their associated uncertainties.  The joint probability is a manner by which to quantify the similarity between the simulated and observed distributions.  This metric was constructed to enable non-parametric comparison of the multidimensional distributions with unknown incompleteness, while accounting for the regions in metallicity space which are only populated by a single distribution.  The first term multiplies the probabilities of pairwise between distributions for each bin in metallicity space, quantifying the degree to which the overlapping regions of the distributions are in agreement.  The first term in brackets weights the amount of overlap in metallicity space by the total probabilities in the overlapping region of the two distributions, while the second term in brackets reduces the value of the metric by weighting the non-overlapping regions by the total probability in these regions.  The final term normalizes the joint probability value by the number of distributions being compared.  The numerical value of this metric has no intrinsic meaning, but it provides a convenient manner of comparison between distributions.  Two perfectly matching distributions will produce a joint probability value of 1, increasing disagreement will produce lower values, including negative values.  The closer the joint probability to 1, the better the agreement between the two distributions.  

The results of the implausibility and joint probability values allow for the evaluation of model parameters in several ways.  One approach is to see if there is a particular set of model parameters that best reproduce the metallicity distribution of a particular element across both of the available observational datasets.  Similarly, it can be seen if there is a consistent set of parameters that best matches several abundances from an individual observational dataset.  Taking these two methods in unison enables two other methods of comparison -- the first is to determine if certain parameter values can be ruled out, and the second is to determine the parameter values that best reproduce the observed metallicity distributions.  

\begin{table*}[htbp]
   \caption{Implausibility and joint probability values when fitting the observed individual [C/Fe]-[Fe/H] and [Mg/Fe]-[Fe/H] distributions at $z=6$.}
   \centering
   \begin{tabular}{@{} llllllllllll@{}} 
& & & \multicolumn{4}{l}{SEGUE}  & \multicolumn{4}{l}{Frebel} \\ 
& & & \multicolumn{2}{l}{C} & \multicolumn{2}{l}{Mg}  &\multicolumn{2}{l}{C}  & \multicolumn{2}{l}{Mg}  \\
IMF &     $\mathcal{E}$   &  $z_{\rm{reion}}$ & Imp  & Joint & Imp  & Joint & Imp  & Joint  & Imp  & Joint   \\
\hline
\hline                              
\rule{0pt}{2ex}Salpeter & 0.04  &   $\leq6$  &  0.66543     &   -34.801     &   0.47764     &   -92.074     &   0.40936     &   -30.786     &   0.40781     &   -65.544       \\
                        &       &   6.5      &  0.66611     &   -33.877     &   0.49025     &   -90.520     &   0.41302     &   -30.089     &   0.42094     &   -63.315       \\    
                        &       &   7        &  0.66306     &   -36.789     &   0.48998     &   -89.357     &   0.41895     &   -32.716     &   0.37755     &   -65.221       \\
                        &       &   8        &  0.66876     &   -34.337     &   0.47427     &   -87.772     &   0.46519     &   -30.056     &   0.57862     &   -62.668       \\  \cline{2-11}
\rule{0pt}{2ex}         & 0.2   &   $\leq6$  &  0.58070     &   -20.459     &   0.35518     &   -89.172     &   0.38601     &   -18.361     &   0.42158     &   -41.291       \\    
                        &       &   6.5      &  0.59219     &   -20.711     &   0.36893     &   -87.351     &   0.38247     &   -19.320     &   0.43138     &   -42.204       \\    
                        &       &   7        &  0.56984     &   -22.462     &   0.37321     &   -85.924     &   0.38771     &   -21.043     &   0.43797     &   -45.974       \\   
                        &       &   8        &  0.63827     &   -23.098     &   0.38780     &   -85.309     &   0.38983     &   -21.543     &   0.44947     &   -46.483       \\  \cline{1-11} 
\rule{0pt}{2ex}Kroupa   & 0.04  &   $\leq6$  &  0.62102     &   -37.026     &   0.45699     &   -93.544     &   0.47112     &   -31.333     &   0.48583     &   -66.337       \\    
                        &       &   6.5      &  0.62239     &   -35.640     &   0.46969     &   -92.333     &   0.46241     &   -30.698     &   0.48671     &   -65.609       \\    
                        &       &   7        &  0.63048     &   -37.698     &   0.46897     &   -90.934     &   0.43870     &   -33.313     &   0.44070     &   -67.088       \\    
                        &       &   8        &  0.72212     &   -33.291     &   0.45183     &   -86.456     &   0.41796     &   -28.987     &   0.47829     &   -61.726       \\  \cline{2-11}
\rule{0pt}{2ex}         & 0.2   &   $\leq6$  &  0.63011     &   -22.976     &   0.35383     &   -89.927     &   0.35720     &   -22.072     &   0.35616     &   -50.513       \\    
                        &       &   6.5      &  0.63580     &   -23.083     &   0.37221     &   -88.632     &   0.34716     &   -22.596     &   0.36389     &   -50.911       \\    
                        &       &   7        &  0.63253     &   -24.976     &   0.37471     &   -87.461     &   0.34359     &   -24.497     &   0.36088     &   -54.992       \\    
                        &       &   8        &  0.63457     &   -24.997     &   0.38655     &   -87.241     &   0.35588     &   -24.072     &   0.36856     &   -53.810       \\  \cline{1-11}
\rule{0pt}{2ex}Chabrier & 0.04  &   $\leq6$  &  0.54636     &   -25.797     &   0.39780     &   -87.988     &   0.34632     &   -23.756     &   0.44308     &   -53.482       \\    
                        &       &   6.5      &  0.54778     &   -25.668     &   0.39778     &   -85.558     &   0.35343     &   -23.591     &   0.43774     &   -52.724       \\    
                        &       &   7        &  0.62486     &   -28.034     &   0.39657     &   -83.948     &   0.43434     &   -25.824     &   0.43712     &   -57.510       \\
                        &       &   8        &  0.61776     &   -26.798     &   0.37787     &   -83.575     &   0.41066     &   -24.759     &   0.45616     &   -55.542       \\  \cline{2-11}
\rule{0pt}{2ex}         & 0.2   &   $\leq6$  &  \bf{0.54602} &   \bf{-13.067} & 0.34978     &   -84.264     &   0.29121     &   \bf{-14.874} &  \bf{0.33708} &  \bf{-30.945}  \\    
                        &       &   6.5      &  0.54788     &   -13.387     &   0.35004     &   -81.828     &   0.29481     &   -15.765     &   0.34166     &   -31.927       \\    
                        &       &   7        &  0.54888     &   -14.568     &   0.34961     &   \bf{-80.059} & \bf{0.28832} &   -17.212     &   0.33926     &   -34.808       \\
                        &       &   8        &  0.63050     &   -15.054     &   \bf{0.34959} &  -81.403     &   0.40477     &   -17.324     &   0.39578     &   -35.978       \\

   \end{tabular}
   \tablecomments{The lowest implausibility and least negative joint probability values at $z=6$ are shown in bold. The simulated elemental-abundance distributions are compared to the observed distributions from both the SEGUE and Frebel datasets.}
   \label{table:ind_imp_and_jp}
\end{table*}

The parameter space of this work is large enough as to prohibit investigating all possible combinations of parameters.  To make this investigation  computationally feasible, a grid of models was first run to $z=10$.  These models were analyzed by comparing the simulated elemental-abundance distribution functions to observations from the SEGUE and Frebel datasets.  From this, several regions of parameter space were ruled out, and further investigation focused on models with parameters that produced particularly promising results.  This initial grid of simulations constrained the chemically-enriched SFE to the intermediate and high values, $0.04$ and $0.2$, and ruled out attenuated values of the LW photon escape fraction, strongly favoring an escape fraction of 1.  This subset of models was further investigated by advancing the simulations to $z=6$ while additionally testing the effect of the redshift of reionization.  Reionization was allowed to commence at redshifts of $8$, $7$, $6.5$, or not at all (which is effectively a redshift of reionization of $z \leq 6$, as that is the redshift where the simulation terminates).  
\begin{figure}[htbp] 
   \centering
   \includegraphics[width=.45\textwidth, clip=true]{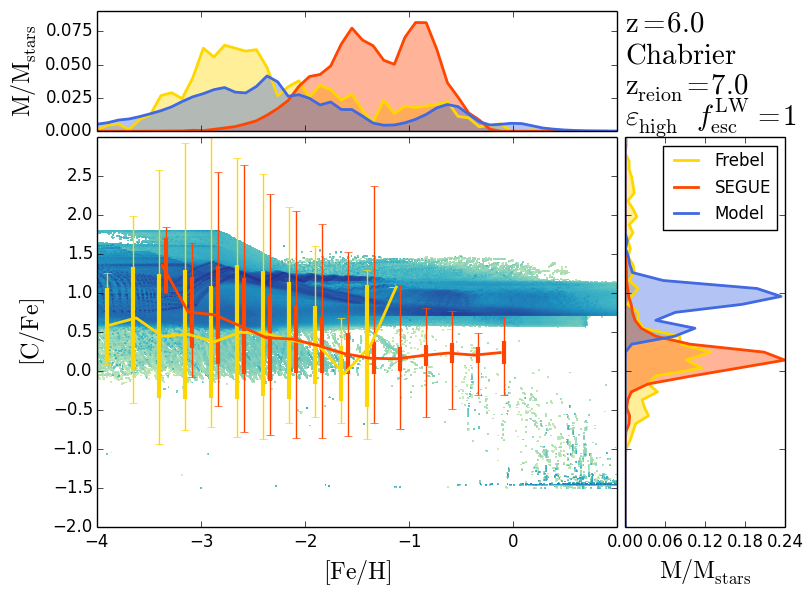} 
   \caption{The [C/Fe]-[Fe/H] distribution at $z=6$ as modeled with a Chabrier IMF, a chemically-enriched SFE of $0.2$, a LW photon escape fraction of 1, and a redshift of reionization of $z_{\rm{reion}}=7$.  The shaded region in the center panel show the distribution of stellar mass in our simulation in [C/Fe]-[Fe/H]  space with the shade reflecting the fraction of stellar mass at that pair of abundances.  Dark blue regions have the largest fraction of the stellar mass, while light blue regions have less stellar mass.  SEGUE data is plotted in red and the Frebel dataset is plotted in yellow.  Observational data is binned in $0.25$ dex increments with the bin mean shown as a continuous line, the $68\%$ confidence interval shown as a thick line, and the maximum and minimum extent of the dataset shown as thin lines.  The top and right histograms show the distributions of stellar mass in either [Fe/H] (top) or [C/Fe] (right).  Simulated data is shown in blue, SEGUE data in red, and data from Frebel in yellow.  This set of parameters maximizes the joint probability for the combined fit of [Mg/Fe] and [C/Fe] in the SEGUE data, and minimizes the implausibility for the same abundances in the Frebel data.}
   \label{fig:chabrier_CFe}
\end{figure}

\begin{figure}[htbp] 
   \centering
   \includegraphics[width=.45\textwidth, clip=true]{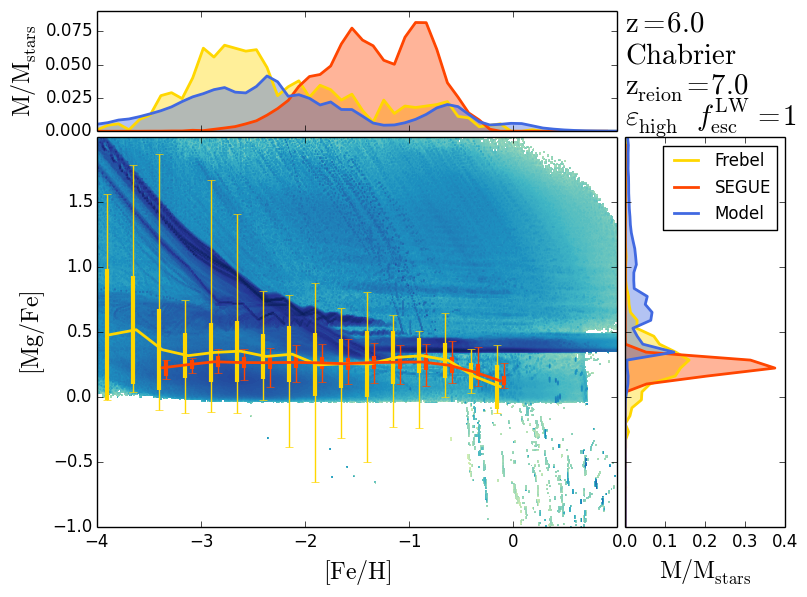} 
   \caption{The [Mg/Fe]-[Fe/H] distribution at $z=6$ for the same model parameters as shown in Figure \ref{fig:chabrier_CFe}, with all coloring and weighting the same as in that figure.  This set of parameters maximizes the joint probability for the combined fit of [Mg/Fe] and [C/Fe] in the SEGUE data and minimizes the implausibility in the Frebel data.}
   \label{fig:chabrier_MgFe}
\end{figure}

The implausibility and joint probability values of various models and the [C/Fe]-[Fe/H] and [Mg/Fe]-[Fe/H] abundance distributions for the SEGUE and Frebel data are shown in Table \ref{table:ind_imp_and_jp}.  Fitting the individual [C/Fe]-[Fe/H] and [Mg/Fe]-[Fe/H] distributions shows a very strong preference for a Chabrier IMF and a SFE of $0.2$. In all cases, for both the SEGUE and Frebel datasets, this pair of parameters minimizes the implausibility and maximizes the joint probability.  Within this pair of parameters, the fits are split between those favoring models with a redshift of reionization of $7$ and those with reionization occurring at $z\le6$.  The variation in the statistical metrics due to changing the redshift of reionization while holding the IMF and SFE fixed is much smaller than the variation when holding fixed the redshift of reionization and changing either the IMF or SFE.  With these metrics we cannot firmly advocate a particular redshift of reionization, but strong constraints can be put on the IMF and SFE.  Varying the SFE while holding the IMF and redshift of reionization constant shows that a SFE of $0.2$ is almost always preferred over an efficiency of $0.04$ in both the implausibility and joint probability metrics.  In comparing parameters sets in which the IMF is the only parameter that varies, the joint probability shows a slight preference for a Salpeter IMF over a Kroupa IMF, the implausibility shows no substantial distinction between the two, but both metrics clearly favor a Chabrier IMF over either of the others.

Evaluating the success of various model parameters with implausibility and joint probability has several caveats that bear consideration.  As can be seen in the top panel of Figures \ref{fig:chabrier_CFe} and \ref{fig:chabrier_MgFe}, the selection functions of the SEGUE and Frebel datasets are very different, with the Frebel dataset being generally composed of more metal-poor stars than the SEGUE dataset.  The discrepancy between the two observed [Fe/H] distributions (which, we note, is purely due to observational selection and not any inherent inconsistency between the datasets) makes a direct and simultaneous comparison to both datasets a difficult prospect.  Additionally, not all model parameter sets produce stellar populations that extend over the same ranges in [X/Fe]-[Fe/H] space.  Some models will have fewer bins with which to compare to observations.  We are restricted in our implausibility analysis to using only bins that have non-zero standard deviations for both model and observational data, and mitigate this difference by normalizing the implausibility values by the number of bins for that parameter set.  

\subsection{Fitting Several Abundances Simultaneously}

\begin{figure}[htbp] 
   \centering
   \includegraphics[width=.45\textwidth, clip=true]{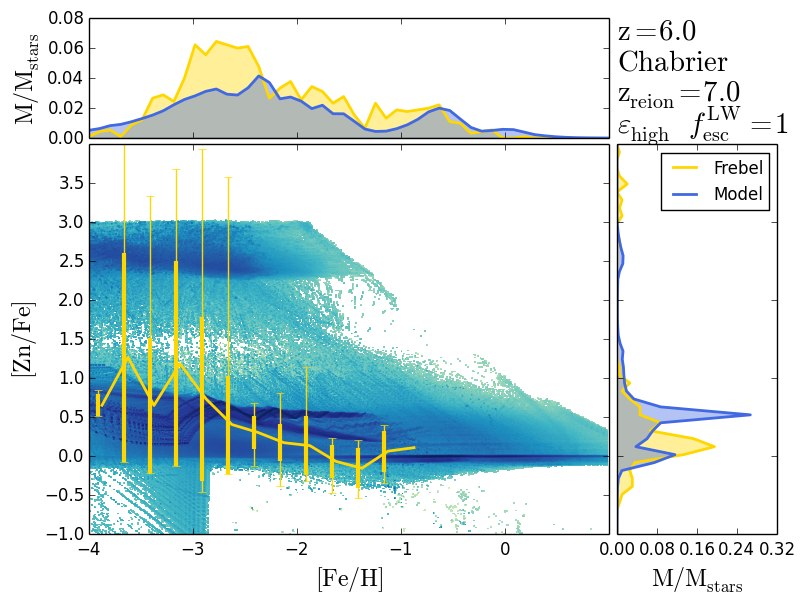} 
   \caption{[Zn/Fe]-[Fe/H] distribution at $z=6$ as produced by a model using a Chabrier IMF, a chemically-enriched SFE of $0.2$, and a redshift of reionization of $7$.  Observational data from Frebel is shown in yellow, binned in $0.25$ dex increments.  The mean in each bin is plotted along with the $68\%$ confidence intervals (thick vertical lines) and maximum and and minimum extent of observed abundance data in each bin.  This model produces the lowest implausibility value when fitting all Frebel abundances simultaneously.}
   \label{fig:chabrier_ZnFe}
\end{figure}

The Frebel dataset contains abundances for 6 elements that are tracked in our model but that are not present in the SEGUE data: N, O, Ca, Ti, Co, and Zn. This set of abundances all have an elemental-abundance distribution with features that are qualitatively different from the observed distributions.  The simulated elemental-abundance distributions have a region at [Fe/H]$<-2.5$ that extends to very low [X/Fe], far below what is observed.  An example can be seen in the [Zn/Fe]-[Fe/H] distribution in Figure \ref{fig:chabrier_ZnFe}.  This feature in the distribution originates from the initial enrichment of a halo by a Population III star where the yields drastically overproduce Fe in relation to the other elements.  The establishes the abundances in this halo at the start of chemically-enriched star formation at levels far less than is observed.

The parameter set shown in Figure \ref{fig:chabrier_ZnFe} is a Chabrier IMF with a high chemically-enriched SFE of $0.2$, a LW photon escape fraction of $1$, and a redshift of reionization of $7$.  The tracks rising from [Zn/Fe]~$<-0.25$ and [Fe/H]~$<-2.8$ to the primary locus in [Zn/Fe]-[Fe/H] space are not seen in the observational data (though we note that this forms a negligible fraction of the total stellar mass -- see the histogram on the right side of the figure).  These tracks originate from the material returned by Population III stars.  This enrichment sets the initial abundances present in the halo at the start of chemically-enriched star formation.  Subsequent generations of star formation and feedback force the abundances in the halo to converge towards the values set by the yields of enriched stellar populations, but the stars created in the early phases of chemically-enriched star formation leave an imprint on the population that is not observed.  This pattern is seen in [N/Fe], [O/Fe], [Ca/Fe], [Ti/Fe], [Co/Fe], and [Zn/Fe].  In all cases, the initial enrichment of the halo from Population III stars leads to the formation of stars with abundances that are not observed, but the abundances in these halos rapidly converge to values that are in better agreement with observations.

\begin{table*}[htbp]
   \caption{Implausibility and joint probability values when simultaneously fitting multiple elemental-abundnaces distributions at $z=6$.}
   \centering
   \begin{tabular}{@{} lllllllll@{}} 

&   &   & \multicolumn{2}{l}{SEGUE C and Mg} &   \multicolumn{2}{l}{Frebel C and Mg} &   \multicolumn{2}{l}{Frebel all}   \\
IMF &   $\mathcal{E}$   &   $z_{\rm{reion}}$    &   Imp    &   Joint   &   Imp    &   Joint   &   Imp    &   Joint   \\
\hline
\hline   
\rule{0pt}{2ex}Salpeter & 0.04  &   $\leq6$ &  0.63812     &   -126.87     &   0.40861     &   -96.330     &   0.45705     &   -83.298       \\
                        &       &   6.5     &  0.64081     &   -124.40     &   0.41683     &   -93.404     &   0.46248     &   -83.215       \\    
                        &       &   7       &  0.63805     &   -126.15     &   0.39871     &   -97.937     &   0.45171     &   -88.469       \\
                        &       &   8       &  0.64041     &   -122.11     &   0.51994     &   -92.724     &   0.48180     &   -84.789       \\  \cline{2-9}
\rule{0pt}{2ex}         & 0.2   &   $\leq6$ &  0.54419     &   -109.63     &   0.40252     &   -59.652     &   0.47584     &   -54.359       \\    
                        &       &   6.5     &  0.55675     &   -108.06     &   0.40511     &   -61.524     &   0.48119     &   -56.499       \\    
                        &       &   7       &  0.53884     &   -108.39     &   0.41098     &   -67.018     &   0.47837     &   -60.883       \\   
                        &       &   8       &  0.59824     &   -108.41     &   0.41745     &   -68.025     &   0.46229     &   -61.408       \\  \cline{1-9} 
\rule{0pt}{2ex}Kroupa   & 0.04  &   $\leq6$ &  0.59569     &   -130.57     &   0.47837     &   -97.670     &   0.47518     &   -88.319       \\    
                        &       &   6.5     &  0.59902     &   -127.97     &   0.47438     &   -96.307     &   0.47820     &   -87.959       \\    
                        &       &   7       &  0.60580     &   -128.63     &   0.43969     &   -100.40     &   0.44923     &   -92.045       \\    
                        &       &   8       &  0.68063     &   -119.75     &   0.44751     &   -90.714     &   0.44359     &   -87.594       \\  \cline{2-9}
\rule{0pt}{2ex}         & 0.2   &   $\leq6$ &  0.58491     &   -112.90     &   0.35671     &   -72.585     &   0.44919     &   -62.975       \\    
                        &       &   6.5     &  0.59372     &   -111.72     &   0.35505     &   -73.507     &   0.45297     &   -65.240       \\    
                        &       &   7       &  0.59168     &   -112.44     &   0.35177     &   -79.489     &   0.44888     &   -69.689       \\    
                        &       &   8       &  0.59483     &   -112.24     &   0.36189     &   -77.882     &   0.43028     &   -69.560       \\  \cline{1-9}
\rule{0pt}{2ex}Chabrier & 0.04  &   $\leq6$ &  0.52509     &   -113.79     &   0.39312     &   -77.238     &   0.46890     &   -68.909       \\    
                        &       &   6.5     &  0.52641     &   -111.23     &   0.39428     &   -76.315     &   0.46747     &   -70.374       \\    
                        &       &   7       &  0.59185     &   -111.98     &   0.43570     &   -83.333     &   0.47859     &   -76.042       \\
                        &       &   8       &  0.58419     &   -110.37     &   0.43262     &   -80.301     &   0.45190     &   -73.087       \\  \cline{2-9}
\rule{0pt}{2ex}         & 0.2   &   $\leq6$ &  \bf{0.51766} &  -97.332     &   0.31258     &   \bf{-45.819} &  0.44074     &   \bf{-40.286}  \\    
                        &       &   6.5     &  0.51934     &   -95.215     &   0.31664     &   -47.692     &   0.43810     &   -42.909       \\    
                        &       &   7       &  0.52004     &   \bf{-94.627} &  \bf{0.31205} &  -52.020     &   \bf{0.43457} &  -46.345       \\
                        &       &   8       &  0.58762     &   -96.457     &   0.40051     &   -53.302     &   0.47000     &   -47.250       \\
   \end{tabular}
   \tablecomments{The lowest implausibility and least negative joint probability values at $z=6$ are shown in bold.  The simulated elemental-abundance distributions are compared to the combined [C/Fe]-[Fe/H] and [Mg/Fe]-[Fe/H] distributions for both the SEGUE and Frebel datasets, as well as to the combined distributions of all elements present in both the simulation and the Frebel data.}
   \label{table:comb_imp_and_jp}
\end{table*}

Multiple abundances within each observational dataset can be fit simultaneously to determine the set of parameters that best reproduces a wide range of the observational data.  The implausibility and joint probability values for various parameter sets resulting from the simultaneous fitting of the observed [C/Fe] and [Mg/Fe] distributions from SEGUE, the [C/Fe] and [Mg/Fe] distributions from Frebel, and all abundances from Frebel are shown in Table \ref{table:comb_imp_and_jp}.  These combinations are best fit with a Chabrier IMF and an elevated chemically-enriched SFE of $0.2$, in agreement with the parameters that provide the best fit to the observed individual [C/Fe]-[Fe/H] and [Mg/Fe]-[Fe/H] abundance distributions. The implausibility and joint probability metrics show a distinct preference for a chemically-enriched SFE of $0.2$, with this value being favored regardless of IMF.  Conversely, varying the IMF while holding the SFE fixed clearly favors a Chabrier IMF, and suggests a slight preference for a Salpeter IMF over a Kroupa IMF.  Similar to the fits of individual abundances, parameter sets with either a redshift of reionization of $7$ or late reionization at $z\leq6$ compare the most favorably with the observed abundance distributions.  If reionization prior to $z=6$ is taken to be mandatory, every fit in which late reionization was favored is supplanted by a one favoring a redshift of reionization of $6.5$.

Extending the model to redshifts below the redshift of reionization has only a small effect on the observed elemental-abundance distribution function.  This can be seen in the very small variations in the implausibility and joint probability values as the redshift of reionization is varied.  Reionization quenches star formation in all but the largest halos, attenuating the formation of new stars and slowing global chemical enrichment.  This is demonstrated in the SFR densities at $z=6$, where the models that include reionization prior to the end of the simulation have SFR densities that are approximately half that of models without reionization.

%% file: discussion.tex
\section{Discussion}
\label{sec:discussion}

\subsection{The Model}
\label{sec:model_improvements}	

This model differs in significant ways from other chemical-evolution models.  Every cosmological halo is tracked independently and evolves as an ensemble of SSPs.  The feedback from these SSPs is a function of the age of each population as well as its metallicity, and takes several forms.  Photodissociating LW radiation is produced by the stellar component of each halo.  Stars and supernovae return enriched material and a quantity of gas available for star formation to the ISM of the halo.  Supernovae provide mechanical feedback that transports gas and enriched material between the central, star-forming region of the halo and the hot, diffuse reservoir in the outer regions, as well as allowing for material to be ejected from the halo entirely and lost to the IGM.  Star formation is a function of the mass of cold, dense gas available in the halo, and the returned enriched material determines the metallicity of the new stars that form out of the gas.

Many of the processes in this model are coupled, resulting in feedback loops that act on both the halo locally and on the global halo population.  The stellar component of all halos contribute to the photodissociating LW background, which determines which halos are capable of forming a Population III star at any given time.  This is coupled to the local state of the chemically-pristine halos through the ability of H$_2$ to self-shield in the presence of LW radiation, and the pristine halo cooling and collapse criteria is a function of the mass of the halo and the current redshift.  Within each halo, the metallicity and density of the gas in the reservoir region determines the cooling time, and along with its mass of gas sets the rate at which gas is exchanged with the central star forming region.  The mass of gas in the central region determines the rate at which new stars are formed.  The metallicity of new stars is determined by the chemical composition of the gas from which they were formed.  The age and metallicity of a stellar population determines the mass of gas and enriched material that will be returned to the ISM of the central region of the halo as well as the number of supernovae that occur during that time step.  These quantities determine the movement of gas and enriched material within the halo, as well as the amount of gas and metals that are ejected from the halo to the IGM.  The multitude of coupled feedback mechanisms make this model substantially more physically representative of the processes occurring in galactic chemical evolution than simpler, one-zone models that treat halos in isolation, or the stellar component of a halo as single SSP.  

This model can be applied to any N-body simulation for which a halo merger tree can be constructed.  Executing this model is much less computationally expensive that running the original simulation with the multiphysics capabilities that would be required to do similar analysis of the chemical evolution.  This model benefits from the ability to post-process a simulation -- it does not need to be implemented at simulation runtime, greatly enhancing its flexibility in regards to the computational facilities that can be used to probe chemical evolution.  Running this model is computationally inexpensive (taking a few hours on a modern desktop computer), enabling the investigation of a wide region of parameter space while using minimal resources.  The modular nature of the code makes it simple to include different sets of stellar yields and update them as new yields become available.  Furthermore,  the treatment of the stellar component of each halo as an ensemble of SSPs allows for the IMF to be varied independently of the yields, and one can introduce new variations such as time- or metallicity-dependent IMFs.  The ability to make statistically significant quantitative comparisons between the results of this model and observational data further bolsters its capabilities, extending its reach beyond what has been achieved in other chemical evolution studies.

\subsection{Comparison to Observations}
\label{sec:comparison_to_obs}
Comparison to observational data is straightforward with this model.  Datasets including substantial numbers of low-metallicity stars, such as the SEGUE spectroscopic dataset \citep{2009AJ....137.4377Y} and the collection of stars with high-resolution spectroscopy compiled by Frebel provide collections of stellar abundances in the ranges of interest to this work.  This comparison does come with a significant caveat:  this model stops at $z=6$, and the available observational datasets are all for low redshift stars in dwarf galaxies and the galactic stellar halo.  The low metallicities of these populations make them prime candidates for use in Galactic Archeology, and suggest that their comparison to the high-redshift chemical evolution is valid.  Bolstering this comparison is the observation that approximately half of all galaxies currently reside in groups, and the Milky Way falls into this category as a member of the Local Group.  At high redshift, the progenitors of the Local Group are statistically average structures, reasonably comparable to the structures in this model.  In addition, previous theoretical work suggests that stars at metallicities below [Fe/H]~$ \simeq -1.5$ almost exclusively come from $z > 6$ \citep{2010ApJ...708.1398T}.  We note that the estimated star formation rate (SFR) density provides an additional method for comparing the model predictions to observations, and that this method is complementary to comparisons of $z=0$ stellar abundances.  High-redshift observations (e.g., \citealt{2011ApJ...737...90B}, \citealt{2014ApJ...786..108O}) are providing constraints on the SFR density at $z\ge10$. 

Using observations of the SFR density at high redshift provides an independent manner of evaluating the ability of a given set of model parameters to reproduce observations.  \citet{2014ApJ...786..108O} report a SFR density of $1.58\times10^{-2}~\mathrm{M_\odot yr^{-1} Mpc^{-3} }$ at $z\sim6$.  The set of parameters favored by abundance fitting, a Chabrier IMF, a chemically-enriched SFE of $0.2$, and a redshift of reionization of $7$, has a SFR density at $z=6$ of $1.23\times10^{-1}~\mathrm{M_\odot yr^{-1} Mpc^{-3} }$, higher than the observed SFR density.  A lower chemically-enriched SFE of $0.04$ provides better agreement with observations.  For example, a Chabrier IMF, with a redshift of reionization of $6.5$, and this lower efficiency produces a SFR density at $z=6$ of $2.55\times10^{-2}~\mathrm{M_\odot yr^{-1} Mpc^{-3} }$.  The tension between our SFR densities and the value reported by \citet{2014ApJ...786..108O} could arise due to differences in the manners in which the SFR density is estimated -- this discrepancy could arise if \citet{2014ApJ...786..108O} are integrating down to a luminosity limit much below ours, or further down the luminosity function but using a shallower slope.

Comparison of our simulated abundance distributions with the SEGUE and Frebel data strongly favor a Chabrier IMF and a chemically-enriched SFE of $0.2$, but did not provide strong constraints on the redshift of reionization.  The most commonly favored reionization parameters were either a redshift of reionization of $7$ or late reionization occurring at $z\leq6$.  Recent work by \citet{2015ApJ...802L..19R} shows that high-redshift star formation is sufficient to produce enough ionizing photons to dominate the reionization of the universe.  The inability of our model to make a strong prediction on a favored redshift of reionization suggests that our model is not particularly sensitive to this parameter.  Inspection of the statistical metrics supports this interpretation.  Variations in the implausibility and joint probability due to different values of the redshift of reionization are much smaller than variations due to changes in the IMF or chemically-enriched SFE.  The preference for parameter sets with a redshift of reionization of $z\le6$ for several observational abundance combinations and statistical metrics suggests that our model may be lacking some physical processes that are necessary to accurately capture this effect.

\begin{figure}[htbp] 
   \centering
   \includegraphics[width=.45\textwidth, clip=true]{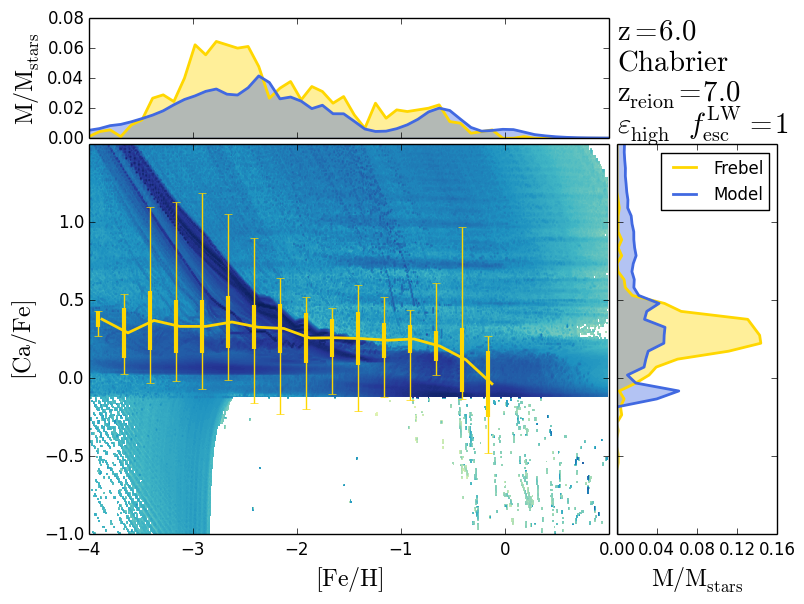} 
   \caption{[Ca/Fe]-[Fe/H] distribution at $z=6$ as produced by a model using a Chabrier IMF, a high chemically-enriched SFE of $0.2$, and a redshift of reionization of $7$.  The majority of observed stars match the simulated metallicity-distribution function, but the qualitative behavior at low values of [Fe/H] differs strongly.  Enrichment from Population III stars drastically under-produces Ca in relation to Fe, establishing initial abundances for chemically-enriched star formation far below those that are observed.}
   \label{fig:chabrier_CaFe}
\end{figure}

\subsection{Population III Stellar Yields}
\label{sec:pop_iii_stellar_yields}
Yields from Population III stars determine the initial abundances in a halo at the start of chemically-enriched star formation.  The current paucity of available Population III stellar yields creates a discrete set of abundances that a halo is enriched to by a Population III star.  These values give rise to the tracks seen in [X/Fe]-[Fe/H] space that converge to the values set by the yields from chemically-enriched stars.  In the cases of [N/Fe], [O/Fe], [Ca/Fe], [Ti/Fe], [Co/Fe], and [Zn/Fe], certain Population III yields set the initial abundances of the halo at levels far below those that are observed.  This manifests as a set of tracks rising in [X/Fe]-[Fe/H] space towards the abundance values that the chemically-enriched yields converge to, as shown in Figure \ref{fig:chabrier_ZnFe} for the [Zn/Fe]-[Fe/H] distribution.  Subsequent generations of chemically-enriched star formation produce distinct tracks in metallicity space as the abundances in the halo converge towards the chemically-enriched yields, but these tracks are not observed in the Galactic stellar halo, indicating that these Population III yields are likely not accurate (or, at the very least, additional stellar yields may be necessary to produce more realistic chemical evolution outcomes).  We note, however, that the mass fraction of stars in our simulations that are contained within these tracks is extremely small, and thus it is entirely possible that we simply have not observed such stars yet.  Another example of this behavior can be seen in Figure \ref{fig:chabrier_CaFe}, which shows the [Ca/Fe]-[Fe/H] distribution at $z=6$.  While our best set of parameters is very successful at matching the observed [Ca/Fe] distribution for the overwhelming majority of the stellar mass, the region with [Fe/H]~$<-2.7$ and [Ca/Fe]~$<-0.2$ is in clear disagreement.  No stars are observed with these combinations of [Ca/Fe] and [Fe/H], but the Population III stellar yields are such that a substantial fraction of chemically-enriched star formation begins at these abundances.  The clear disagreement between the observed and simulated yields strongly suggests that the simulated Population III stellar yields do not reflect reality in some important way.

\subsection{Possible Constraints from Hydrodynamical Simulations}
\label{sec:constraints_from_sims}
This work highlights several areas where model parameters can be constrained with results from high-resolution hydrodynamical simulations.  Such simulations can help to elucidate the gas distribution in dark matter halos -- for example, the fraction of gas that is cold and dense, as well as the radius in which this dense gas is contained, are aspects of this model that can be refined with the help of hydrodynamic simulations.  Simulations of supernova feedback in simulated high-redshift galaxies \citep[e.g.,][]{2012ApJ...745...50W} can also guide the refinement of the models of metal and gas transport in the halos, as can simulations investigating the nature of accretion and halo mergers.  High-resolution simulations using more sophisticated star formation algorithms can aid in the constraint of the chemically-enriched SFE, one of the primary free parameters in the model of chemically-enriched star formation used in this work.  Furthermore, in our models we implicitly ignore any interactions between halos other than the H$_2$-photodissociating LW background and halo mergers, which includes such effects as halo cross-pollution and ionizing regions around galaxies.  While analytic estimates suggest that it is reasonable to ignore these effects, physics-rich cosmological simulations can test the validity of these assumptions.

\subsection{Limitations and Future Work}
\label{sec:limitations}

A complete discussion of the limitations of the star formation model is presented in Paper I, but a brief recap of several salient points is given here.  Ionizing radiation is treated as a uniform metagalactic background rather than being modeled as a local interaction between halos.  Local interactions may be particularly important for quenching star formation in small halos, through as was shown in Paper I, these halos host very little star formation and contribute a negligible amount to the total SFR and chemical enrichment.  Additionally, material that is ejected from a halo does not interact with any surrounding halos.   This material could conceivably be accreted by a nearby halo, but this is very unlikely, as chemically-enriched material ejected by a supernova will only extend to a radius of $\sim1$ kpc in $10^5-10^7$ years \citep{2003ApJ...596L.135B,2008ApJ...682...49W}, and will have a negligible effect on nearby satellite minihalos \citep{2010ApJ...712..101W}.  The role of ionizing radiation and SNe ejecta will be investigated in future work, but can be neglected here as the impact would not be global, and would have a minimal effect on the overall abundance-ratio distributions.

Our simulations stop at $z=6$, which presents a potential caveat for comparison with observational data gathered at $z=0$.  This is a reasonable, as [Fe/H] $< -1.5$ stars almost all form at $z > 6$ \citep{2010ApJ...708.1398T}, enabling us to compare our simulations to the low-metallicity stellar halo.  While this enables comparison between observations and our work, it is imperfect.  In looking at the relationship between stellar age and metallicity at $z=0$, \citet{2010ApJ...717..542K} and \citet{2010MNRAS.401L...5S} find that a substantial fraction of stars with [Fe/H]~$>-1.5$ have ages of less than $12.5~$Gyr, making their formation redshift lower than the redshift at which our simulations end.  We can more confidently compare the results of our model at $z=6$ to observations at $z=0$ by restricting our analysis to stars with [Fe/H]$<-2.5$, the vast majority of would have formed during the redshifts simulated in our work \citep{2006ApJ...638..585F,2007MNRAS.381..647S,2010ApJ...708.1398T}.  Larger simulations volumes capable of running to $z=0$ and the addition of more complete, self-consistent stellar yields data will help address these issues.  Future work will extend our models to $z=0$ using high-resolution cosmological N-body simulations.  Extending our simulations to $z=0$ will enable comparisons with more and larger observational datasets, both those that are currently available (e.g., SEGUE) and those that will become available in the near future such as LAMOST \citep{2012ARA&A..50..251I}, APOGEE \citep{2008AN....329.1018A}, Gaia-ESO \citep{2012Msngr.147...25G}, GALAH \citep{2012ASPC..458..421Z}, and RAVE \citep{2003ASPC..298..381S}.

We have incomplete coverage of stellar masses and metallicities in the available yields, and what coverage we do have is not self-consistent.  Different models (i.e., simulation codes) with different fundamental assumptions go into separately producing AGB, SNII, and SNIa yields since we employ abundances from different authors who have different codes, and who often only do near-solar and primordial compositions.  The lack of reliable low-metallicity yields (below approximately 0.1 Z$_\odot$) is understandable -- stellar evolution models are calibrated using local stars that are typically close to Solar metallicity -- but presents challenges to those who wish to self-consistently model galactic chemical evolution.  We are working with collaborators that are experts in stellar evolution, and are in the process of generating a self-consistent grid of yields using the MESA code \citep{2011ApJS..192....3P, 2013ApJS..208....4P} that will span the necessary range of masses and metallicities, and which will be used in future work.

Finally, future work will include better methods of comparing data to observations, and of analyzing the relationship between model inputs and observations.  We are currently using simple metrics such as joint probability and a generic measure of implausibility, and small grids of models.  However, more sophisticated techniques, such as Gaussian Multiprocess emulation coupled with Markov Chain Monte Carlo tools and ANOVA decomposition of our models, have been developed by our group for other purposes \citep{2012ApJ...760..112G, 2014ApJ...787...20G}, and will soon be applied to chemical evolution.

%% file: conclusions.tex
\section{Summary and Conclusions}
\label{sec:conclusions}

This work presents a new semi-analytical model of chemical evolution that can be applied to cosmological N-body simulations of large populations of high-redshift galaxies, and which can be compared directly to abundance measurements of the Milky Way stellar halo.  Our model assumes that star formation occurring over a short period of time results in ``simple stellar populations'' (SSPs) having uniform metallicity and identical star formation time, and uses publicly-available abundance tables from simulations of AGB and Type Ia and Type II supernovae to calculate the feedback of metal-enriched gas and energy into the halo's interstellar medium.  A single halo can be composed of many of these SSPs with a range of ages, and its overall output at any point in time is the sum of the SSP outputs at that time.  We consider a range of model inputs, including a halo's chemically-enriched star-formation efficiency, the escape fraction of H$_2$-photodissociating LW photons, the choice of IMF fitting function, and the choice of nucleosynthetic abundances that are put into the SSPs.  We compare our model outputs to two observational datasets -- the SEGUE stellar sample \citep{2009AJ....137.4377Y}, and also a sample of several hundred metal-poor stars with detailed abundances compiled by \citet{2010AN....331..474F}.  Our primary results are as follows:

\begin{enumerate}
\item  The model parameters that best reproduce the observed abundance-ratio distributions are a Chabrier IMF, a chemically-enriched star formation efficiency of $0.2$ (similar to a galactic gas depletion time of approximately half a Gyr \citep{2011ApJ...730L..13B}) and a LW photon escape fraction of $1$.  

\item  The redshift of reionization is much more weakly constrained than the IMF and chemically-enriched star-formation efficiency.  This is likely because by a redshift of $8$ -- the earliest that we allow for the onset of reionization -- the majority of star formation is occurring in halos that are large enough to not be strongly affected by reionization.

\item Other features in the simulated abundance-ratio distributions suggest inaccuracies in the Population III stellar yields.  Abundances in halos at the start of chemically enriched star formation are set by the yields of Population III stars, and in the cases of N, O, Ca, Ti, Co, and Zn, are drastically less than the observed abundances, indicating that these elements are being underproduced in relation to Fe in the yields calculations.  

\end{enumerate}

More broadly, we have introduced a new model for the chemical evolution of galaxy populations that can be coupled to large N-body simulations, and thus has the capability to self-consistently provide both spatial and temporal information about star formation, chemical evolution, and other quantities relating to Milky Way progenitors.  This model can be both qualitatively and quantitatively compared to the observed abundance distributions in large stellar surveys, and we have designed our model outputs so that it will be straightforward to couple the model to sophisticated statistical tools \citep[e.g.,][]{2012ApJ...760..112G, 2013MNRAS.436.3602G}.  These tools will enable detailed, quantitative comparison of models to both current and future observations of Milky Way stellar populations, and are particularly useful when dealing with multiple large datasets (such as combinations of many different observables).

%% file: acknowledgments.tex
\section{Acknowledgments}
\label{sec:acknowledgements}

BDC and BWO were  supported by the National Science Foundation under
Grant No. PHY-1430152 (JINA Center for the Evolution of the Elements).  
BWO was supported
by the National Aeronautics and Space Administration through grant
NNX12AC98G and Hubble Theory Grant HST-AR-13261.01-A.  He was also
supported in part by the sabbatical visitor program at the Michigan
Institute for Research in Astrophysics (MIRA) at the University of
Michigan in Ann Arbor, and gratefully acknowledges their hospitality.  TCB
acknowledges 
partial support for this work from grants PHY08-22648; Physics Frontier 
Center/Joint Institute or Nuclear Astrophysics (JINA), and PHY 14-30152; 
Physics Frontier Center/JINA Center for the Evolution of the Elements (JINA-CEE),
awarded by the US National Science Foundation.  The
simulations presented here were performed and analyzed on the NICS
Kraken and Nautilus supercomputing resources under XSEDE allocations
TG-AST090040 , and the semi-analytical models were performed using
MSU's High Performance Computing Center.  We thank Britton Smith and
Facundo Gomez for helpful discussions during the course of preparing
this manuscript, Carolyn Peruta for sharing code and for useful
discussions,
and an anonymous referee for comments which substantially improved
the quality of the manuscript.  \texttt{Enzo} and \texttt{yt} are developed by a
large number of independent researchers from numerous institutions
around the world.  Their commitment to open science has helped make
this work possible.

%% file: appendix.tex
\appendix

\section{Stellar Yields}
\label{app:yields}
Creation of the feedback tables was accomplished by convolving the yields of a grid of stellar evolution models with a stellar initial mass function (IMF) to create chemical and kinetic feedback tables for an integrated stellar population.  These tables encapsulate information regarding feedback as a function of the time since formation of the stellar population, and assume that all stars are formed in accordance with the adopted IMF and the metallicity-dependent stellar lifetime.  These tables can be applied to a single simple stellar population of a given age, and in the case of a halo that has experienced star formation in its past the tables can be used to determine the total amount of chemical and kinetic feedback by applying them separately to each stellar age bin in the halo.  Separate tables were created for AGB and super-AGB stars, as well as for Type Ia and Type II supernovae.  These tables were created using several metallicities.  The modular nature of the yield tables allow for the investigation of the impact of individual sources, and the success or failure of yields for different stellar metallicities.  Additionally, new yields can easily be integrated into the model and tested independently of the functionality of the rest of the code. 

\subsection{Yields of Asymptotic Giant-Branch Stars}
\label{sec:agbyield}
Intermediate-mass stars are presumed to be the main producers of heavy s-process nuclides, and also contribute substantially to the yields of several other nuclides, most notably carbon and nitrogen, during their AGB phase \citep{2007A&A...476..893S}.  \citet{2007PASA...24..103K} and \citet{2010MNRAS.403.1413K} calculated detailed stellar models and post-processed nucleosynthetic data to produce AGB yields. Their models cover a range in mass from 1.0 M$_{\odot}$ to 6 M$_{\odot}$ and compositions Z = 0.02, 0.008, 0.004, and Z = 0.0001, where Z is the metal mass fraction.  All models were evolved from the zero-age main sequence to near the tip of the thermally-pulsing AGB phase.  \citet{2010MNRAS.403.1413K} used an updated set of proton- and $\alpha$-capture rates, and assumed scaled-solar abundances for low-metallicity models, rather than adopting the initial abundances of the Small and Large Magellanic Clouds as was done in \citet{2007PASA...24..103K}.

Super-AGB stars are defined by a specific mass range between the minimum mass for carbon ignition and the mass limit above which the star ignites neon at its center, and evolves through all nuclear burning stages up to an iron-core collapse supernovae \citep{2007A&A...476..893S}.  The mass range of super-AGB stars varies with metallicity, with a lower limit between 7.5 and 9 M$_\odot$, and an upper limit of approximately 11 M$_\odot$ \citep{2007A&A...476..893S,2010A&A...512A..10S}.  In \citet{2010A&A...512A..10S}, yields are computed in a post-processing step, with initial mass ranges focused on covering the computationally demanding thermal pulses. The \citet{2010A&A...512A..10S} model computes the evolution of convective-zone abundances, with instantaneous mixing of chemical species, and allows for different nuclear reaction rates and uncertainties within various spatial and temporal regions of the star.  Due to these considerations, we use the yields of \citet{2010MNRAS.403.1413K} for AGB stars, in conjunction with yields from \citet{2010A&A...512A..10S} for super-AGB stars.  

\subsection{Yields of Type II Supernovae}
\label{sec:typeII}

Stars with a zero age main sequence mass of 8-40 M$_\odot$ are expected to end their lives as Type II supernovae (SNII) \citep{2003ApJ...591..288H}.  The rate of SNII is in turn determined by calculating the main sequence lifetime of stars in this mass range using the mass-age relation of \citet{1996A&A...315..105R}.  These lifetimes, used in conjunction with an IMF, enable us to calculate the SNII rate as a function of time for a stellar population of a given mass.

\citet{2006ApJ...653.1145K} calculated yields for stars of metallicity $Z $~= 0, 0.001, 0.004, 0.02 and masses in the range 13 - 40 M$_{\odot}$. Final yields are tuned to produce 0.07 M$_{\odot}$ of ejected iron. \citet{1998A&A...334..505P} calculated a set of yields ranging for stars from 6-120 M$_\odot$ with metallicities from $Z=0.0004-0.05$ undergoing explosive nucleosynthesis.  In \citet{1998A&A...334..505P}, supernovae are triggered by electron captures on heavy nuclei, photo-dissociation of iron into $\alpha$-particles, and rapid neutralization of collapsing material.  Our model utilizes a combination of yields from \citet{2006ApJ...653.1145K} for stellar masses 13-40 M$_\odot$, \citet{1998A&A...334..505P} for masses 40-120 M$_\odot$, and rates derived from \citet{1996A&A...315..105R}.  

\subsection{Yields of Type Ia Supernovae}
\label{sec:typeIa}
Thermonuclear supernovae are important contributors to the chemical enrichment of the ISM, primarily with iron-peak nuclei and some intermediate-mass nuclei. SNIa are usually modeled as explosions of white dwarfs that have approached the Chandrasekhar limit ($M_{ch}\sim1.39M_{\odot}$) through accretion from a companion in a binary system \citep{1982ApJ...253..798N}.  Progenitor models are classified as either single degenerate, in which a white dwarf grows in mass due to accretion from an evolving binary companion, and double degenerate, in which two C-O white dwarfs merge.  Our model uses yields from \citet{1999ApJS..125..439I}, which are based on the progenitor model of \citet{1997NuPhA.621..467N} and employs a single-degenerate scenario.  The SNIa rates are from \citet{2009ApJ...707.1466K}, which again uses a single-degenerate scenario for modeling the SNIa progenitor.